Observation of Quantum Capacitance of individual single walled carbon nanotubes

Junfeng Dai[1], Jun Li[1], Hualing Zeng[1], Xiaodong Cui[1,2*]
1.Department of Physics, the University of Hong Kong, Hong Kong, China
2. Department of Chemistry, the University of Hong Kong, Hong Kong, China
Corresponding email: xdcui@hku.hk

**Abstract**: We report a measurement on quantum capacitance of individual semiconducting and small band gap SWNTs. The observed quantum capacitance is remarkably smaller than that originating from density of states and it implies a strong electron correlation in SWNTs.

In the classical regime, capacitance describes the object's storage capability of electric charges and it is solely determined by electrostatic force, eventually by the geometry and the medium's dielectric constant. As the object's spatial dimensions shrink to nanometer scale where quantum effects have to be taken account, the states accommodating charges are numbered and the electric charges are regulated by the number of available states as well as electrostatic force: The Pauli exclude principle prohibits electrons/holes from occupying the same states. Therefore capacitance arising from finite density of states is significant in quantum regime, especially in nanometer scale. Besides the modification stemming from finite density of states, electron-electron interaction may give a further correction on the quantum capacitance under some circumstance. As the potential difference and the number of electric charges are simply additive, these capacitances consequently contribute to the total capacitance C in series, $C_{tot}^{-1} = C_g^{-1} + C_Q^{-1}$ where $C_g$ and $C_Q$ are respectively the geometric and quantum capacitance. For these nanometer scale samples and low dimensional objects such as

2DEGs nanostructures, [1, 2], nanotubes, nanowires,[3-5]nanojunctions [2, 6] and quantum dots, [7, 8] their quantum capacitance is comparable to the classical one and therefore the quantum capacitance measurement as a probe to fundamental aspects of electronic properties is possible and has been discussed theoretically and experimentally since late 80s. Recently McEuen group firstly reported their observation of quantum capacitance of a top-gated individual carbon nanotube.[4] In this letter, we present our experimental measurements on capacitance of individual microscopically long single walled carbon nanotubes (SWNT) at the temperature range of 77-300K. The measured capacitance is remarkably less than the simulated one from density of states based on tight-binding calculation and the difference may result from the strong electron correlation.

Individual macroscopically long and diluted SWNTs were grown on a degenerately doped $Si/SiO_2$ (oxide thickness=300nm) substrate using the chemical vapor deposition method. The catalyst was prepared by selectively dipping the diluted solution of *FeCl₃* on the silicon substrate and then by being reduced in the furnace under an $Ar/H_2$(400 sccm/50 sccm) at 900°C for 20 minutes. Individual SWNTs were formed on the substrates in ethanol vapour with the same gas mixture at 900°C for a hour. The diameters of SWNTs measured by resonant Raman scattering [9] and AFM are in range of 1.3-2.5nm. Electric contacts to SWNTs were made by standard optical lithography, Ti/Al metal deposition on top of as-prepared SWNTs on substrates and finally lift off. The spacing between metal electrodes varies from 10μm to several hundreds of micrometers. For these semiconducting SWNTs contacting to two or more electrodes, field effect transistors are formed with the degenerated silicon substrate as a back gate electrode. Electric characterization of these long channel (up to 1000μm) FETs shows standard p type or ambipolar behaviors. It implies that structural discontinuity or high electric barrier due to defects are rare along SWNTs up to 1000μm. The capacitance measurement was carried out by adding a small sinusoidal bias of one kilohertz on the back gate and monitoring the displace current through the SWNT connected electrodes. To get rid of contribution from the electrodes' capacitance and the parasite capacitance of the background we feed both displace currents from a SWNT connected electrode and from a suspended electrode of the identical geometry to a balanced amplifier, as sketch in Figure 1. So the current arising from the capacitance of electrodes was canceled out and the MOS-C like behavior of silicon substrates was eliminated as well. Finally the balanced signal was fed to a lock-in amplifier which extracts the signal amplitude and phase information, and the capacitance and resistance were analyzed. All electric measurements were carried out in vacuum condition (~$10^{-2}$Pa) at the temperature range of 77-300K. The geometric structure of the specific SWNTs was *in situ* identified by resonant micro-Raman scattering where 532nm and 633nm lasers were selected as excitation sources. [10]

Figure 2 shows the measured capacitance of a single SWNT (16,8) as a function of Vg at various temperatures. The curves exhibit a zero baseline at low gate bias, corresponding to the "off" state of standard SWNT-FETs. In this case, the Schottky barrier between electrodes and the SWNT prevents electrons/holes from charging and meanwhile the Fermi level lies amid the SWNT band gap and therefore no states available to accommodate charges. While Vg increases towards either negative or positive, the Schottky barrier at electrode contacts becomes thinner and lower and the Fermi level approaches to the first sub-band and therefore electrons/holes start to charge the SWNT. Consequently the capacitance exhibits remarkable increase within the gate voltage limits of ±10V until a plateau of 8pF appears at $V_g$<-6V, where the first sub-band is fully involved. Noticeably the capacitance at "on" state keeps unchanged at the temperature range of 77-160K.

Comparing with the signal from the capacitance (the X component of the Lock-in outout), the quadrature component (Y component) exhibits negligible change in the whole gate voltage range at 77K. It implies the resistance of the whole system is negligibly small compared to *$1/\omega C$ ($\omega=1$KHz)*, which is consistent with independent electric characterization, and therefore SWNTs are under equilibrium at the testing frequency even though electrons propagate diffusively along the SWNT. Since the capacitance is a measure of charging effect, whether electron transport is ballistic or diffusive does not affect the capacitance measurement as long as time scale is long enough for charging. For the presented SWNT of 330μm long measured by SEM, it is however questionable that there exist defects and local disorders which act as local barriers to low energy electrons/holes. This question was examined by studying the capacitance dependence on the AC bias amplitude. Figure 2(c) shows that the "on" state capacitance measured at various AC amplitudes keeps constant in the amplitude range of 0.8-2V. While the AC amplitude is lower than 0.7V, the capacitance drops significantly due to existence of barriers either against propagating electrons/holes or against charge injection and consequently the effective length of SWNT shrinks. Once the AC amplitude is high than 0.8V, barriers vanish and the effective SWNT length expands to the whole physical length of 330μm. The fact that the capacitance at "on" state keeps unchanged at various temperature also supports this scenario.

Another possible factor affecting our measurement on intrinsic SWNT capacitance lies in the interface capacitance between metal electrode and the SWNT, which can be treated as a serial capacitor connecting to the SWNT. To estimate this capacitance, we performed a direct measurement on several SWNT-FET structures, as sketched in Figure 3(a). A small low frequency (*1K-10KHz*) sinusoidal bias was fed into the drain electrode; the current through the source electrode was monitored by a lock-in amplifier. The system capacitance and the resistance could be extracted from the current amplitude and phase information according to LRC circuit model. Figure 3(b) shows the conductance and phase through a SWNT-FET with a source-drain channel of 10μm as a function of $V_g$. Note that the phase signal at low gate voltage corresponding to "off" state is overwhelmed by electronic noise. Since the inductance of SWNTs is estimated at tens nH/μm (Kinetic inductance),[11] $R >> \omega L$, its contribution to the impedance is negligible at low frequency. Based on the simplified RC circuit analysis, we can obtain the capacitance around 10pF at maximum at "on" state, where the Schottky barrier becomes thinnest. Measurements over several SWNT FET structures with various channel length show similar results on the interface capacitance. As the interface capacitance (~pF) is three orders of magnitude greater than that of the whole capacitance(~fF) as shown in Figure 2, its contribution to the total capacitance is negligible.

To extract the quantum capacitance of single SWNT (16,8), we assume that (1) $C_{total}^{-1} = C_g^{-1} + C_Q^{-1}$; and (2) $C_g = \frac{2\pi\varepsilon L}{\cosh^{-1}(2h/d)} \approx \frac{2\pi\varepsilon L}{\ln(4h/d)}$ where $d$ and $L$ are the

SWNT diameter of 1.67nm and length of 330μm, *h=300nm*(further confirmed with SEM imaging) is the thickness of oxide layer and the dielectric permittivity $\varepsilon= 3.9\varepsilon_0$ for $SiO_2$; (3) the whole capacitance is sum of segments in parallel. Accordingly we obtained the quantum capacitance $C_Q$ for the SWNT (16,8) and the quantum capacitance rises from zero at low gate bias to 82aF/μm at Fermi level sitting at its first sub-band. It is however significantly smaller than the simulated value arising from density of states calculated from the zone folded tight-binding simulation, as shown in Figure 2(d),[12] where we define

$$C_{DOS}(u) = \frac{dQ(u)}{du} = e\int_0^\infty \frac{\partial f(V-u)}{\partial u} * DOS(V)dV.$$

Figure 4(a) shows the measured capacitance as a function of Vg of a single small band gap SWNT which has geometric index (n,m) satisfying n-m=3Xinteger. [13, 14] This kind of SWNTs are expected to be metallic by tight binding calculation, but are in fact semiconducting with small band gap when measured. This results from curvature effect that couples $P_\pi$ and σ orbits to open a small gap. The capacitance measurements shown in Figure 4(a) tell the band gap disappears above or at 160K. Following the similar treatment as described earlier, one can extract the quantum capacitance of 10.3aF/μm for this SWNT upon the Fermi level lying in the first sub-band. The Raman signal could not be obtained and the exact geometric index was unknown as its optical transition does not resonate with the excitation laser lines. But generally quantum capacitance of metallic SWNTs can be estimated as follows: The energy for accommodating one electron is supposed to be equal to that of electron at Fermi level, say

$$\frac{e^2}{C_{DOS}} = \hbar V_F K = \hbar V_F \frac{2\pi}{L}$$

then one gets the quantum capacitance per unit length (considering the spin degeneracy)

$$\frac{C_{DOS}}{L} = \frac{2e^2}{hV_F} \approx 96\,aF/\mu m$$

Where $V_F$ is the Fermi velocity and is set at $8\times10^5$ m/s for metallic SWNTs.

In general, the quantum capacitance obtained in the experiments is remarkably smaller than that from DOS simulated with tight-binding calculation. It implies that a single particle model cannot well describe the situation here and strong electron correlation must be considered. Theoretically a one dimensional interacting electron gas system is described by Tomonaga-Luttinger liquid instead of Fermi liquid, where electron-electron interaction qualitatively instead of perturbatively modifies electric properties due to the reduced dimensionality. Its validation on carbon nanotubes, a quasi 1D system was proved by independent experiments. [15, 16] In Luttinger liquid model, charges propagate in form of charge density wave with a velocity that is different from Fermi velocity and is determined by the strength of the interaction, saying $V_C = \frac{V_F}{g}$, where $V_C$ denotes the charge velocity, and g is a dimensionless parameter describing the electron-electron interaction strength and therefore the compressibility of the liquid: g>1 for attractive interaction; g=1 for non-interaction; and g<1 for repulsive interaction. As the electron correlation affect the compressibility of the liquid, the velocity and consequently the capacitance (a naïve analog is a classical acoustic wave, where capacitance C→density ρ, wave velocity $V \propto 1/\sqrt{\rho}$) , the factor g can be extracted by

$$g = \sqrt{\frac{C_Q}{C_{DOS}}}.$$ [17] Taking this model in mind, we estimate at 77K g in range of 0.25-0.3 for SWNT (16,8) and g around 0.32 for the small band gap SWNT in Figure 3 when the Fermi level lying at the first sub-band. The values agree with previously reported experiments.[4, 16, 18, 19]

In conclusion, we obtained the quantum capacitance of individual semiconducting and small band gap SWNTs. The observed quantum capacitance is remarkably smaller than the capacitance originating from density of states and it implies a strong electron correlation in SWNTs.

**Acknowledgement:** We thank Guanhua Chen, Jian Wang, Xudong Xiao and Liwei Chen for helpful discussion. The work was supported by Research Grant Council of Hong Kong under Grant No.: HKU 7019/06P.

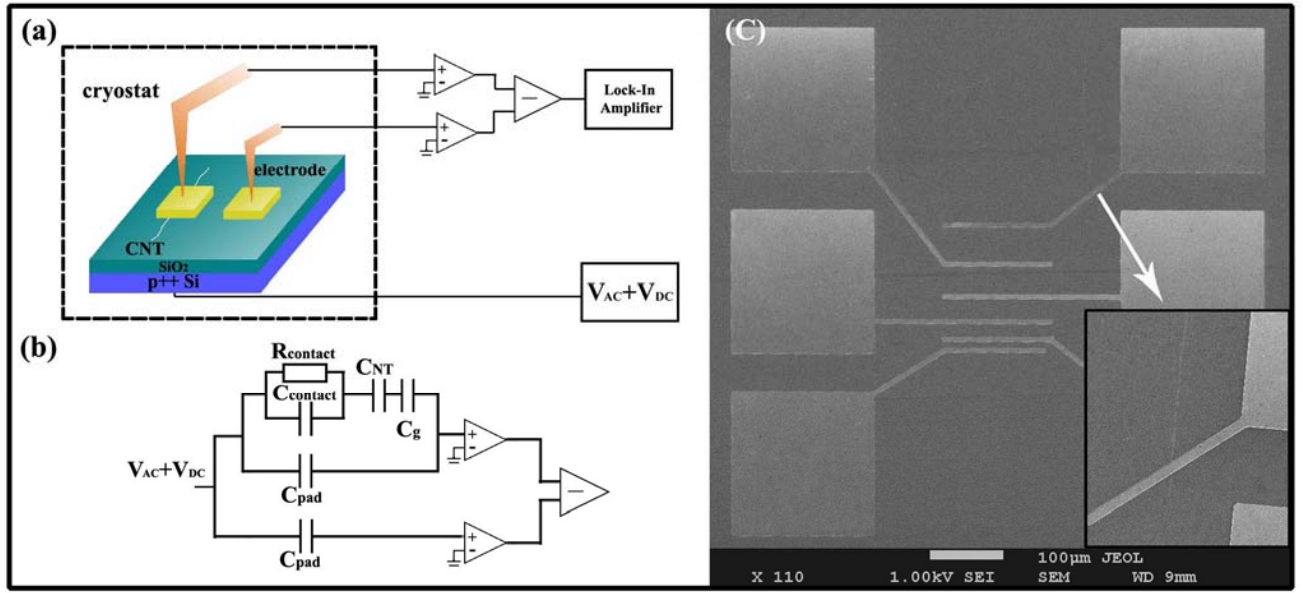

Figure 1. (a) Schematic of the capacitance measurement. The displace currents through electrode pads connected with and without SWNTs are fed into a differential current amplifier so that the contribution from the electrodes are cancelled out. Then the signal is collected by a Lock-in amplifier and the in-phase (X component) and quadrature component (Y component) are simultaneously obtained. Based on the equivalent circuit model (b), the capacitance can be derived. (c) A SEM image of devices on silicon substrates. Inset is a magnified zoom in of an electrode connected with an individual SWNT (white).

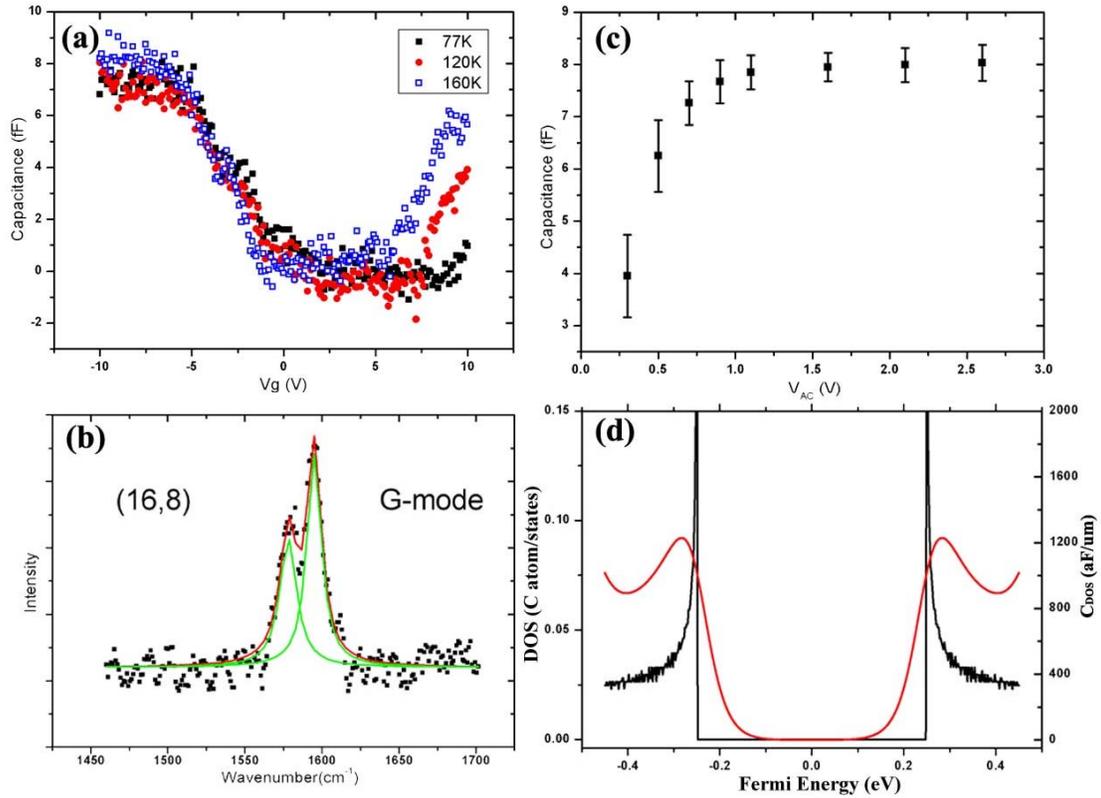

Figure 2. (a) The measured capacitance of a SWNT of 330μm as a function of gate voltage at various temperatures. (b) The G-mode Raman scattering of the SWNT with the excitation laser of 532nm. Radial Breathing Mode (RBM) cannot be obtained with lasers of 532nm and 632.8nm owing to off-resonance. Two Lorentzian fittings (in green) are used to find the G- and G+ position. The diameter is obtained by the relation of

$d = \sqrt{\dfrac{\omega_G^+ - \omega_G^-}{C}}$ , where C=47.7cm-1 for semiconducting SWNTs.[10] With the aid of Kataura plot, we identified its geometric index as (16,8).[10]    (c) The measured capacitance as a function of AC bias amplitude at 77K. Error bars give the signal range averaged from many runs. (d) A simulated density of states of a SWNT (16,8) and the corresponding capacitance arising from density of states as a function of Fermi energy, based on a zone folded tight-binding calculation.

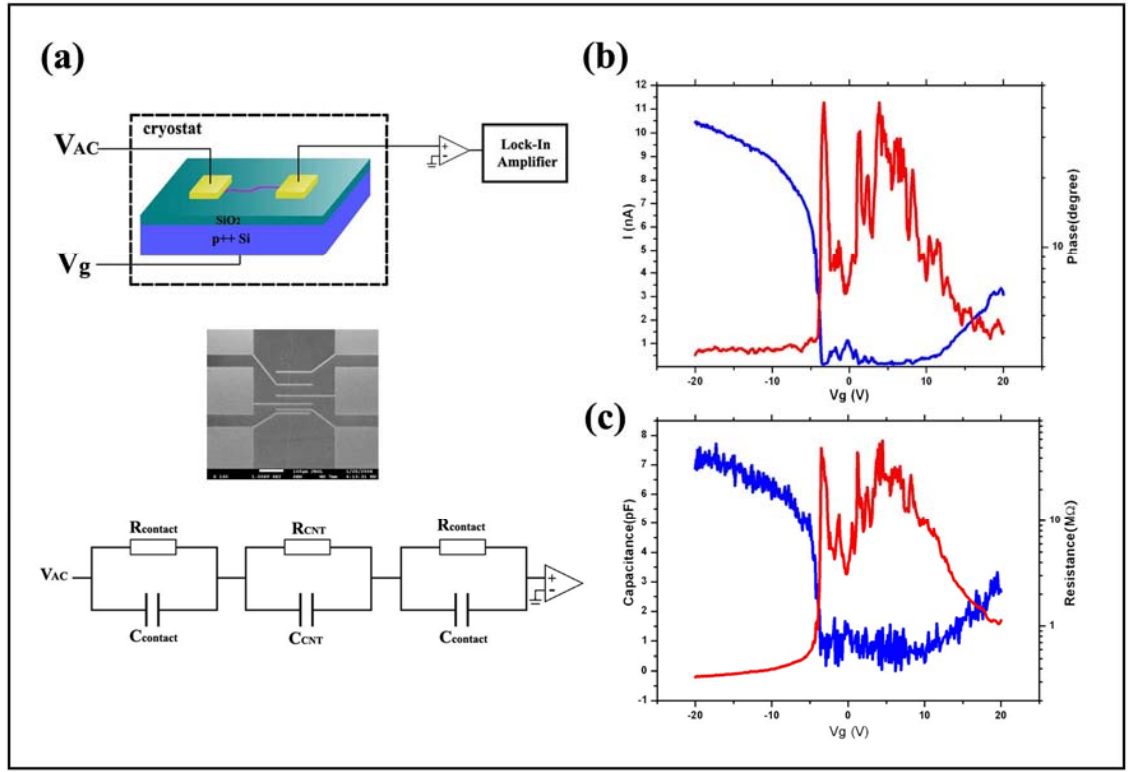

Figure 3. (a) Schematic of the capacitance between electrodes and SWNT measurements. A small 1KHz sinusoidal bias (<10mV) is fed into the drain electrode and the current through the source electrode is monitored by a lock-in amplifier. Both the current amplitude and phase are analyzed to derive the capacitance and system resistance, based on the equivalent circuit model. (b) The measured current amplitude (blue) and phase(red). (c) The calculated capacitance (blue) and resistance (red) of the SWNT as a function of back-gate voltage.

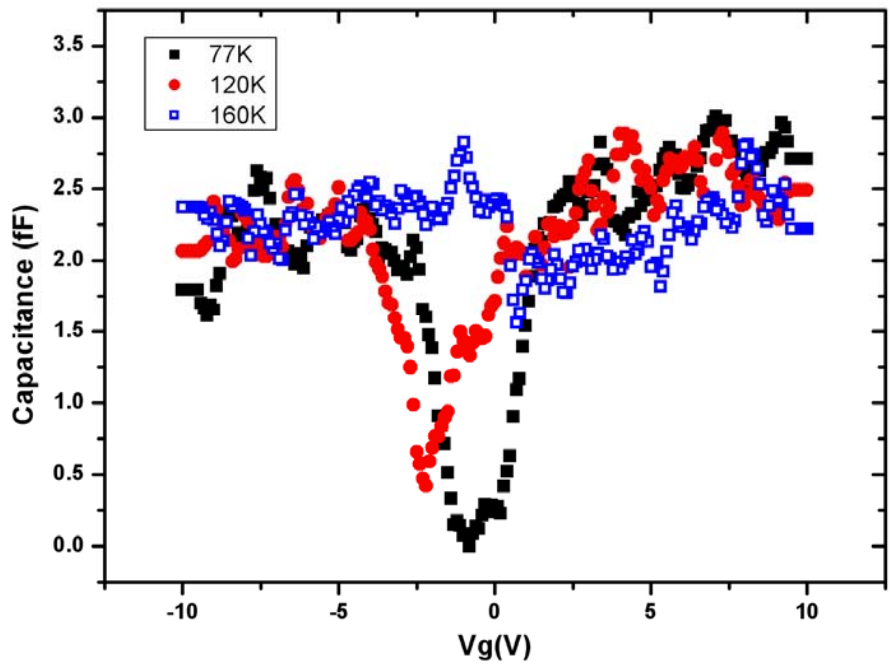

Figure 4. The measured capacitance of a small band gap SWNT at temperature of 77K, 120K and 160K.